\documentclass{llncs}
\usepackage{amsfonts}
\usepackage{upgreek}
\usepackage{stmaryrd}
\usepackage{amssymb}
\usepackage{paralist}
\usepackage{amsmath}
\usepackage[boxed,ruled]{algorithm2e}
\usepackage{listings}
\usepackage[dvips]{graphicx}
\begin{document}

\title{Secure Information Flow\\ by Model Checking Pushdown System\protect\footnote{This work was partially supported by the National
Natural Science Foundation of China under Grant No. 60773163.}}

\author{Cong Sun \and Liyong Tang \and Zhong Chen
\institute{ Institute of Software, School of EECS, Peking
University, China\and Key Laboratory of High Confidence Software
Technologies, Ministry of Education, China}
\email{\{suncong,tly,chen\}@infosec.pku.edu.cn}} \maketitle

\begin{abstract}
We propose an approach on model checking information flow for
imperative language with procedures. We characterize our model with
pushdown system, which has a stack of unbounded length that
naturally models the execution of procedural programs. Because the
type-based static analysis is sometimes too conservative and rejects
safe program as ill-typed, we take a semantic-based approach by
self-composing symbolic pushdown system and specifying
noninterference with LTL formula. Then we verify this LTL-expressed
property via model checker Moped. Except for overcoming the
conservative characteristic of type-based approach, our motivation
also includes the insufficient state of arts on precise information
flow analysis under inter-procedural setting. To remedy the
inefficiency of model checking compared with type system, we propose
both compact form and contracted form of self-composition. According
to our experimental results, they can greatly increase the
efficiency of realistic verification. Our method provides
flexibility on separating program abstraction from noninterference
verification, thus could be expected to use on different programming
languages.
\end{abstract}
\section{\label{sec:intro}Introduction}

Noninterference is a standard criterion to formalize secure
information flow. This property was first introduced by Goguen and
Meseguer\cite{DBLP:conf/sp/GoguenM82} for multi-level computing
system and applied to programming language via semantic model. It
commonly means any two runs of program starting in two
indistinguishable states yield two indistinguishable final states.
That is to say, for any pair of runs the difference on secret input
are unobservable via public output.

Typical information flow analyses are studied using security type
system
\cite{DBLP:journals/jcs/VolpanoIS96}\cite{DBLP:conf/tapsoft/VolpanoS97}\cite{DBLP:conf/popl/SmithV98}\cite{DBLP:conf/popl/HeintzeR98}\cite{mcps:sabefeld}.
These type systems guarantee well-typed programs do not leak any
secret information. But these type-based approaches are considered
overly conservative and sometimes reject safe programs, such as
$l\mathrel{\mathop:}= h\cdot 0$. Take the following program as
another example,

\begin{minipage}[t]{4.8in}
\begin{lstlisting}[language=C,basicstyle=\footnotesize]
                if (l) y:=h; else skip;
                if (!l) x:=y; else skip;
\end{lstlisting}
\end{minipage}
Suppose $l$ and $x$ are \textsf{low}, while $h$ and $y$ are
\textsf{high}, standard type
system\cite{DBLP:journals/jcs/VolpanoIS96} rejects this program
over-restrictively by reporting a flow from $h$ to $x$.

To remedy this problem, semantic-based
approaches\cite{DBLP:journals/scp/JoshiL00}\cite{DBLP:journals/lisp/SabelfeldS01}
are developed, especially a recent popular approach,
self-composition
\cite{DBLP:conf/csfw/BartheDR04}\cite{DBLP:conf/sas/TerauchiA05},
which composes the original program with a variable-renamed copy to
avoid considering correlative executions as pair. Model checking is
an important method to verify properties of semantic-based program
model. It is well-known as fully-automated and mature
tool-supported. Though argued against the high complexity, model
checking has been used to check noninterference, combined with
abstract interpretation\cite{DBLP:journals/fuin/FrancescoST03},
self-composition\cite{DBLP:conf/sas/TerauchiA05}\cite{DBLP:conf/pldi/UnnoKY06},
and type-based approach\cite{DBLP:conf/pldi/UnnoKY06}.

Lots of existing work get into the spectrum of discussing
noninterference over more complex program constructs. Procedure is
considered as a common feature of realistic imperative languages. In
some applications we need to verify code from different sources
together. Also there may be context-sensitive procedure calls in
program. Noninterference verification of these programs involves
inter-procedural information flow. Consider the program below

\begin{minipage}[t]{4.8in}
\hspace*{1cm}
\begin{minipage}[lt]{3cm}
\begin{lstlisting}[language=C,basicstyle=\footnotesize]
 int h,l;
 ...
 func(h,l);
 ...
\end{lstlisting}
\end{minipage}
\begin{minipage}[r]{4.8cm}
\begin{lstlisting}[language=C,basicstyle=\footnotesize]
void func(int a, int &b) {
    int c:=0;
    while (a>0) { c++; a--; }
    b:=c;
}
\end{lstlisting}
\end{minipage}
\end{minipage}
after \texttt{func} returns, the \textsf{high} variable $h$ has
indirectly passed to the \textsf{low} variable $l$. Suppose the
global variables are only observable after $\textsf{func}$ returns,
inter-procedural analysis is required. Volpano and
Smith\cite{DBLP:conf/tapsoft/VolpanoS97} has developed a
non-standard type system to check information flow of an imperative
language with procedures. But for the more precise
approaches\cite{DBLP:conf/pldi/UnnoKY06}\cite{DBLP:conf/sas/TerauchiA05}\cite{DBLP:journals/fuin/FrancescoST03},
procedure is not considered.\\
\indent Some work treats information flow with program logic in
procedural setting. Beringer et al.\cite{DBLP:conf/csfw/BeringerH07}
adapt type-based approach to Hoare-like logic and introduce an
auxiliary binary formulae to encode noninterference according to the
principle of self-composition. Their procedure model is only
restricted to parameterless form. Amtoft et
al.\cite{DBLP:conf/popl/AmtoftBB06} propose a Hoare-like logic to
analyze inter-procedural information flow with independence
assertion for object-oriented language. Method call in
object-oriented language introduces additional impreciseness because
of pointer aliasing on concrete location. Their procedure form is
different from procedure of imperative languages and the memory
model is much more complex.\\
\indent Hammer et al.\cite{Hammer06informationflow} investigate an
approach for information flow control based on a well-known program
slicing technique, \emph{program dependence graph} (PDG).
Inter-procedural information flow control can be achieved by
computing summary graph and constructing \emph{system dependence
graph} (SDG). They introduce path
condition\cite{DBLP:journals/tosem/SneltingRK06}\cite{DBLP:conf/sas/Snelting96}
to improve the preciseness. Since PDG has been applied to handle
realistic programs in C and Java, their approach seems quite
general. But in their approach impreciseness comes from the
construction of PDGs/SDGs. Programs should be translated into
\emph{static single form} (SSA) and the size of path condition
should be reduced. This method is also much more expensive than
type-based approaches and conservative because we cannot exclude
safe program if some PC(y,x)$\equiv$true.\\
\indent The relative insufficient study on inter-procedural
information flow in state of arts motivates us to propose a general
inter-procedural framework for more precise information flow
analyses than the type-based approaches. Our model is based on
symbolic pushdown system\cite{mcps:schwoon}. A pushdown system is a
transition system with a stack of unbounded length in states. This
unbounded stack provides a natural way to model the execution of
program with procedure. The state of pushdown system includes
control locations and stack symbols. Control locations are used to
store global variables. Stack symbols contain both control points
and local variables of procedure. This semantic model has been
proved sound to our language and can be directly expressed as the
input of model checker Moped\cite{mcps:moped}.\\
\indent The verification process consists of the following phases.
First a symbolic pushdown system is derived from core-language
program. Then we self-compose the derived symbolic pushdown system
and express noninterference with LTL formula as the inputs of Moped.
Verification is then performed by Moped. If noninterference is
violated by the program model, we can get a single counterexample.
This witness trace can help us find out which \textsf{high} variable
causes the insecure information flow, thus be useful for secure
program development. We have experimentally proved that our method
is more precise than ordinary type-based analysis on
inter-procedural information flow. Considering the relative high
complexity of model checking, we have developed two derived forms of
ordinary self-composition, called \emph{compact self-composition}
and \emph{contracted self-composition}. The experiments indicate
great efficiency improvement with these derived forms compared with
ordinary self-composition. Another advantage of our approach is the
flexibility unveiled by self-composing symbolic pushdown system
instead of high-level language. Unlike other work did
self-composition directly on high-level
languages\cite{DBLP:conf/pldi/UnnoKY06}\cite{DBLP:conf/sas/TerauchiA05},
our approach separates the possible abstractions of high-level
languages from noninterference verification. Although we use a
simple imperative language with procedure for simplicity, we believe
it possible to apply our approach to different high-level
programming
languages by abstracting them into pushdown systems.\\
\indent The rest of our paper is structured as follows. Section
\ref{sec:modeling} sketches the simple imperative language and
presents symbolic pushdown system derivation. Section
\ref{sec:prop_spec} proposes the algorithms of both ordinary form
and derived forms of self-composition for composing pushdown system,
and introduces how to specify noninterference with LTL. We report
our experiments in Section \ref{sec:expr} and possible discussions
in Section \ref{sec:discuss}.

\section{\label{sec:modeling}Program Modeling}
\subsection{\label{subsec:language}Language Syntax and Semantics}

The syntax of our core language with first-order procedure is given
in Fig.~\ref{fig:syntex}. $l$ is memory location. The \textsf{in}
parameter is local while the \textsf{out} parameter represents
reference to global variable. The big-step operational semantics are
presented in Fig.~\ref{fig:semantics}. $\mu,\lambda\subseteq
\mathbb{L}\times \mathbb{V}$ are heap and stack respectively, where
$\mathbb{L}$ is the domain of memory location and $\mathbb{V}$ is
the domain of value. $dom(\mu)\cap dom(\lambda)=\emptyset$.
$\lambda\uplus[l\mathrel{\mathop:}= v]$ extends $\lambda$ with a new
$l$ assigned with $v$. BINDVAR stores $v$ to new location $l$ of
$\lambda$ and refers each free occurrence of $x$ in $S$ to $l$.
Therefore the scope of $x$ is $S$, and $l$ is deallocated after the
execution of $S$. We assume all substitutions capture-avoiding and
procedure $main$ non-recursive.

\begin{figure}[t]
\parbox[t]{4.8in}{\footnotesize
$\mathrm{v \mathrel{\mathop:\mathop:}= c\ \arrowvert\ \textbf{true}\ \arrowvert\ \textbf{false}}$\hfill(value)\\
$\mathrm{e \mathrel{\mathop:\mathop:}= x\ \arrowvert\ v\ \arrowvert\
l\ \arrowvert\ e_1\circ
e_2\ \arrowvert\ \textbf{proc}(\textsf{in}\ x_1, \textsf{out}\ x_2)S}$\hspace*{\fill}(expression)\\
$\mathrm{S \mathrel{\mathop:\mathop:}=\ \textbf{skip}\ \arrowvert\ e
\mathrel{\mathop:}= e'\ \arrowvert\ \textbf{if}\ e\ \textbf{then}\
S_1\ \textbf{else}\ S_2}\
\mathrm{\arrowvert\ \textbf{while}\ e\ \textbf{do}\ S\ \arrowvert\ S_1;S_2}\\
\hspace*{0.8cm}\mathrm{\arrowvert\ \textbf{letvar}\
x\mathrel{\mathop:}= e\ \textbf{in}\ S\ \arrowvert\ e(e_1,
e_2)}$\hspace*{\fill}(statement)}
\caption{\label{fig:syntex}Language Syntax}
\end{figure}

\begin{figure*}[t]
\begin{minipage}[t]{6.8in}\scriptsize
$\mathrm{\frac{\displaystyle\ }{\displaystyle
(\mu,\lambda,v)\downarrow v}}$ \quad(VAR)\qquad
$\mathrm{\frac{\displaystyle (\mu,\lambda,e_1)\downarrow v_1\qquad
(\mu,\lambda,e_2)\downarrow v_2}{\displaystyle (\mu,\lambda,e_1
\circ e_2)\downarrow v_1\circ v_2}}$ \quad(BINOP)

$\mathrm{\frac{\displaystyle\ }{\displaystyle (\mu,\lambda,l)
\downarrow \mu(l)}\ l\in dom(\mu)}$\quad (HEAPLOC)\qquad
$\mathrm{\frac{\displaystyle\ }{\displaystyle (\mu,\lambda,l)
\downarrow \lambda(l)}\ l\in dom(\lambda)}$\quad (STACKLOC)

$\mathrm{\frac{\displaystyle(\mu,\lambda,S_1)\downarrow
(\mu',\lambda')\qquad (\mu',\lambda',S_2)\downarrow
(\mu'',\lambda'')}{\displaystyle (\mu,\lambda,S_1;S_2)\downarrow
(\mu'',\lambda'')}}$\quad(SEQUENCE)

$\mathrm{\frac{\displaystyle(\mu,\lambda,e)\downarrow
\textbf{true}\qquad (\mu,\lambda,S_1)\downarrow
(\mu',\lambda')}{\displaystyle(\mu,\lambda,\textbf{if}\ e\
\textbf{then}\ S_1\ \textbf{else}\ S_2)\downarrow
(\mu',\lambda')}}$\quad(IF-T)\qquad $\mathrm{\frac{\displaystyle
(\mu,\lambda,e)\downarrow \textbf{false}\qquad
(\mu,\lambda,S_2)\downarrow (\mu',\lambda')}{\displaystyle
(\mu,\lambda,\textbf{if}\ e\ \textbf{then}\ S_1\ \textbf{else}\
S_2)\downarrow (\mu',\lambda')}}$\quad(IF-F)

$\mathrm{\frac{\displaystyle(\mu,\lambda,e)\downarrow
v}{\displaystyle(\mu,\lambda,l\mathrel{\mathop:}= e)\downarrow
(\mu[l\mathrel{\mathop:}= v],\lambda)}\quad l\in
dom(\mu)}$\quad(UPDATE-HEAP)

$\mathrm{\frac{\displaystyle(\mu,\lambda,e)\downarrow
v}{\displaystyle(\mu,\lambda,l\mathrel{\mathop:}= e)\downarrow
(\mu,\lambda[l\mathrel{\mathop:}= v])}\quad l\in
dom(\lambda)}$\quad(UPDATE-STACK)

$\mathrm{\frac{\displaystyle}{\displaystyle(\mu,\lambda,\textbf{skip})\downarrow(\mu,\lambda)}}$\quad
(SKIP)\qquad $\mathrm{\frac{\displaystyle (\mu,\lambda,e)\downarrow
\textbf{false}}{\displaystyle (\mu,\lambda,\textbf{while}\ e\
\textbf{do}\ S)\downarrow (\mu,\lambda)}}$\quad(WHILE-F)

$\mathrm{\frac{\displaystyle (\mu,\lambda,e)\downarrow
\textbf{true}\qquad (\mu,\lambda,S)\downarrow
(\mu',\lambda')\qquad(\mu',\lambda',\textbf{while}\ e\ \textbf{do}\
S)\downarrow (\mu'',\lambda'')}{\displaystyle
(\mu,\lambda,\textbf{while}\ e\ \textbf{do}\ S)\downarrow
(\mu'',\lambda'')}}$\quad(WHILE-T)

$\mathrm{\frac{\displaystyle(\mu,\lambda,e)\downarrow v\qquad
(\mu,\lambda\uplus[l\mathrel{\mathop:}= v],[l/x]S)\downarrow
(\mu',\lambda')}{\displaystyle (\mu,\lambda,\textbf{letvar}\
x\mathrel{\mathop:}= e\ \textbf{in}\ S)\downarrow
(\mu',\lambda'\setminus \{l\})}\qquad l\notin dom(\mu)\ and\ l
\notin dom(\lambda)}$\quad(BINDVAR)

$\mathrm{\frac{\displaystyle (\mu,\lambda,e)\downarrow v\qquad
(\mu,\lambda\uplus[l'\mathrel{\mathop:}=
v],[l'/x_1][l/x_2]S)\downarrow (\mu',\lambda')}{\displaystyle
(\mu,\lambda,(\textbf{proc}(\textsf{in}\ x_1,\textsf{out}\
x_2)S)(e,l))\downarrow (\mu',\lambda'\setminus \{l'\})}\qquad
l'\notin dom(\mu)\ and\ l'\notin dom(\lambda)}$\quad(CALL)
\end{minipage}
\caption{\label{fig:semantics}Induction Rules}
\end{figure*}

\subsection{\label{subsec:pds}Pushdown System}

\begin{definition}[Pushdown System]
A \emph{pushdown system} is a triple
$\mathcal{P}=(P,\Gamma,\Delta)$. $P$ is a finite set of control
location, $\Gamma$ is stack alphabet, and $\Delta\subseteq (P\times
\Gamma)\times (P\times {\Gamma}^*)$ is a finite set of pushdown
rules. The pushdown rule has a form of $\langle
p,\gamma\rangle\hookrightarrow  \langle p',w\rangle$ to represent
the relation $((p,\gamma),(p',w))$, where $p,p'\in P, \gamma\in
\Gamma, w\in \Gamma^*$. A \emph{configuration} of $\mathcal {P}$ is
a pair $\langle p,w\rangle$, where $p\in P$ and $w\in {\Gamma}^*$.
$w$ stands for a snapshot of stack content. Suppose the
configuration set of $\mathcal {P}$ is denoted by $Conf(\mathcal
{P})$. We have a transition relation $\twoheadrightarrow \subseteq
Conf(\mathcal {P})\times Conf(\mathcal {P})$ defined by the set of
pushdown rules $\Delta$ as follows: If $\langle p,\gamma\rangle
\hookrightarrow \langle p',w\rangle$, then $\langle p,\gamma
w'\rangle \twoheadrightarrow \langle p',ww'\rangle$ for all $w'\in
\Gamma^*$. Let $c_0$ be the initial configuration. We have a
transition system corresponding to the pushdown system as ${\mathcal
{I}}_{\mathcal {P}}=(Conf(\mathcal {P}),\twoheadrightarrow ,c_0).$
\end{definition}

Each pushdown rule gives a pattern of program execution step.
Without loss of generality, we assume $\mid w\mid \leq 2$ in that
any pushdown system can be put into a normal form($\mid w\mid \leq
2$) with linear size increase\cite{mcps:schwoon}.

\begin{definition}[Symbolic Pushdown System]
A \emph{symbolic pushdown system} is a pushdown system with form
$(P_0\times G,\Gamma_0\times L,\Delta,C_0)$. $P_0$ is the set of
symbolic control locations, and $\Gamma_0$ is stack alphabet. $G$
and $L$ are respectively the domain of the control locations and
stack symbols. The pushdown rules have a form $\langle
(p,g)(\gamma,l)\rangle \hookrightarrow \langle
(p',g'),(\gamma_1,l_1),\ldots,(\gamma_n,l_n)\rangle $, where
$p,p'\in P_0, \gamma,\gamma_1,\ldots,\gamma_n\in \Gamma_0$. The
corresponding symbolic form separates the symbolic part from a
relation of valuations:\\
\hspace*{\fill}$\langle p,\gamma\rangle \hookrightarrow \langle
p',\gamma_1,\ldots,\gamma_n\rangle\qquad(g,l,g',l_1,\ldots,l_n)\in R$\hspace*{\fill}\\
where $R\subseteq (G\times L)\times (G\times L^n)$ is a relation.
Here we can also generally suppose $n\leqslant 2$. $\Delta$ is the
set of symbolic pushdown rules. $C_0=(\{p_0\}\times G)\times
(\{y_0\}\times L)$ is the set of initial configurations.
\end{definition}

The global variables are encoded into control locations and the
local variables are encoded into the stack alphabet. Since we
specify the control points of each procedure with stack symbols, and
the global and local variables are considered as their value in $R$,
$P_0\times G$ can be simplified to $G$ and the symbolic pushdown
rule is of form $\langle \gamma\rangle \hookrightarrow \langle
\gamma_1,\ldots,\gamma_n\rangle,\ (g,l,g',l_1,\ldots,l_n)\in R$.

\subsection{\label{subsec:derivation}Pushdown Rules Derivation}

\begin{figure}[b]
\begin{minipage}[t]{4.8in}\footnotesize
$\mathrm{F(\textbf{skip}, p)=\emptyset}$\qquad
$\mathrm{F(x\mathrel{\mathop:}= e, p) =} \left\{
\begin{array}{ll} \mathrm{\langle p,x\rangle} &
\mathrm{if\ x\notin dom(\mu)}\\
\mathrm{\emptyset} & \mathrm{if\ x\in dom(\mu)}\end{array} \right.$

$\mathrm{F(S_1;S_2, p)=F(S_1, p)\cup F(S_2, p)}$\hspace*{\fill}
$\mathrm{F(\textbf{letvar}\ x\mathrel{\mathop:}= e\ \textbf{in}\ S,
p)=F(S, p)\cup\{\langle p,x\rangle \}}$

$\mathrm{F(\textbf{while}\ e\ \textbf{do}\ S,
p)=F(S,p)}$\hspace*{\fill} $\mathrm{F(\textbf{if}\ e\ \textbf{then}\
S_1\ \textbf{else}\ S_2, p)=F(S_1, p)\cup F(S_2,p)}$

$\mathrm{F((\textbf{proc}(\textsf{in}\ x_1,\textsf{out}\
x_2)S)(e,l), p)=\{\langle p',x_1\rangle \}\cup F([l/x_2]S,
p')}$,\hspace*{\fill} $\mathrm{p'}$ is the tag of
$\mathrm{\textbf{proc}}$
\end{minipage}
\caption{\label{fig:unification}Unifying Local Variables}
\end{figure}
\begin{figure*}[t]\footnotesize
\begin{minipage}[t]{4.8in}
$\mathrm{\Phi(\textbf{skip},n_i,n_j,p,R)=\{\langle
(\rho_i)(n_i,\eta_i(p))\rangle \hookrightarrow \langle
(\rho_i)(n_j,\eta_i(p))\rangle \mid R\} }$\hfill(\texttt{P-SKIP})

$\mathrm{\Phi(x\mathrel{\mathop:}=
e,n_i,n_j,p,R)=\{\langle(\rho_i)(n_i,\eta_i(p))\rangle
\hookrightarrow \langle(\rho_j)(n_j,\eta_j(p))\rangle \mid
R'\}}$\hspace*{\fill}(\texttt{P-UPDATE})\\
\hspace*{\fill}$\left\{ \begin{array}{llllll}
\mathrm{R'=R\cup\{\rho_j[x]=e\}} &\wedge &
\mathrm{\rho_j=\rho_i[x\mathrel{\mathop:}= e]} &\wedge &
\mathrm{\eta_j(p)=\eta_i(p)}, & \mathrm{if\ x\in \rho}\\
\mathrm{R'=R\cup \{\eta_j(p)[x]=e\}} &\wedge &
\mathrm{\rho_j=\rho_i} &\wedge &
\mathrm{\eta_j(p)=\eta_i(p)[x\mathrel{\mathop:}= e]}, & \mathrm{if\
x\in \eta(p)}\end{array}\right.$\hspace*{\fill}

$\mathrm{\Phi(S_1;S_2,n_i,n_j,p,R)=\Phi(S_1,n_i,n_k,p,R)\cup
\Phi(S_2,n_k,n_j,p,R)}$\hfill(\texttt{P-SEQ})

$\mathrm{\Phi(\textbf{if}\ e\ \textbf{then}\ S_1\ \textbf{else}\
S_2,n_i,n_j,p,R)=\{\langle (\rho_i)(n_i,\eta_i(p))\rangle
\hookrightarrow \langle
(\rho_i)(n_k,\eta_i(p))\rangle \mid R\cup \{e\}\}}\\
\hspace*{\fill}\mathrm{\cup \{\langle (\rho_i)(n_i,\eta_i(p))\rangle \hookrightarrow \langle (\rho_i)(n_l,\eta_i(p))\rangle \mid R\cup \{!e\}\}}\\
\hspace*{\fill}\mathrm{\cup \Phi(S_1,n_k,n_j,p,R\cup \{e\})\cup \Phi(S_2,n_l,n_j,p,R\cup \{!e\})}$\\
\hspace*{\fill}(\texttt{P-BRANCH})

$\mathrm{\Phi(\textbf{while}\ e\ \textbf{do}\
S,n_i,n_j,p,R)=\{\langle (\rho_i)(n_i,\eta_i(p))\rangle
\hookrightarrow \langle (\rho_i)(n_j,\eta_i(p))\rangle \mid R\cup
\{!e\}\}\cup}\\
\hspace*{\fill}\mathrm{\{\langle (\rho_i)(n_i,\eta_i(p))\rangle
\hookrightarrow \langle (\rho_i)(n_q,\eta_i(p))\rangle \mid R\cup
\{e\}\}\cup \Phi(S,n_q,n_i,p,R\cup \{e\})}$\quad (\texttt{P-LOOP})

$\mathrm{\Phi(\textbf{letvar}\ x\mathrel{\mathop:}= e\ \textbf{in}\ S,n_i,n_j,p,R)=}\\
\mathrm{\{\langle (\rho_i)(n_i,\eta_i(p))\rangle \hookrightarrow
\langle (\rho_i)(n_k,\eta_i(p)[x\mathrel{\mathop:}=
e])\rangle \mid R\cup \{\eta_k(p)[x]=e\}\}\cup \Phi(S,n_k,n_j,p,R)}$\\
\hspace*{\fill}(\texttt{P-BINDVAR})

$\mathrm{\Phi((\textbf{proc}(\textsf{in}\ x_1,\textsf{out}\ x_2)S)(e,l),n_i,n_j,p,R)=}\\
\hspace*{\fill}\mathrm{\{\langle (\rho_i)(n_i,\eta_i(p))\rangle
\hookrightarrow \langle
(\rho_i)(n_k,\eta_k(p')[x_1\mathrel{\mathop:}=
e])(n_j,\eta_i(p))\rangle \mid R\cup \{\eta_k(p')[x_1]=e\}\}}\\
\hspace*{\fill}\mathrm{\cup \Phi([l/x_2]S,n_k,n_q,p',\emptyset)\cup
\{\langle (\rho_q)(n_q,\eta_q(p'))\rangle \hookrightarrow \langle
(\rho_q)(\varepsilon)\rangle \mid R\}}$, $\mathrm{p'}$ is the tag of $\mathrm{\textbf{proc}}$\\
\hspace*{\fill}(\texttt{P-PROC})
\end{minipage}\caption{\label{fig:derivation}Derivation Rules for Pushdown System}
\end{figure*}
\begin{figure}[t]\footnotesize
\begin{minipage}[t]{4.8in}
$\mathrm{\{\langle n_1\rangle \hookrightarrow \langle n_3,n_2\rangle
\mid\ \{h'=h,l'=l,x_1'=h,c'=c\}\}\cup}$ $\mathrm{\{\langle
n_4\rangle \hookrightarrow \langle
\varepsilon\rangle \mid \{h'=h,l'=l\} \}\cup}$\\
$\mathrm{\{\langle n_3\rangle \hookrightarrow \langle n_5\rangle
\mid \{h'=h,l'=l,x_1'=x_1,c'=0\}\}\cup}$\\
$\mathrm{\{\langle n_5\rangle \hookrightarrow \langle n_7\rangle
\mid \{x_1>0,h'=h,l'=l,x_1'=x_1,c'=c\}\}\cup}$\\
$\mathrm{\{\langle n_7\rangle \hookrightarrow \langle n_8\rangle
\mid \{x_1>0,h'=h,l'=l,x_1'=x_1,c'=c+1\}\}\cup}$\\
$\mathrm{\{\langle n_8\rangle \hookrightarrow \langle n_5\rangle
\mid \{x_1>0,h'=h,l'=l,x_1'=x_1-1,c'=c\}\}\cup}$\\
$\mathrm{\{\langle n_5\rangle \hookrightarrow \langle n_6\rangle
\mid \{x_1\leq 0,h'=h,l'=l,x_1'=x_1,c'=c\}\}\cup}$\\
$\mathrm{\{\langle n_6\rangle \hookrightarrow \langle n_4\rangle
\mid \{h'=h,l'=c,x_1'=x_1,c'=c\}\}\cup}$ $\mathrm{\{\langle
n_2\rangle \hookrightarrow \langle \varepsilon\rangle \mid
\{h'=h,l'=l\} \}}$
\end{minipage}
\caption{\label{fig:example1}Symbolic Pushdown Rules}
\end{figure}

Suppose local variable $x$ of procedure $p$ is uniquely denoted by
$\langle p,x\rangle \in \mathbb{P}\times \mathbb{Q}$, where
$\mathbb{P}$ is the domain of procedure tag and $\mathbb{Q}$ is the
domain of local variable name. Fig.~\ref{fig:unification} gives
static analysis $\mathrm{F}$ to get the unified local variable set
$\theta\subseteq \mathbb{P}\times \mathbb{Q}$. Partial function
$\mathrm{\eta: \mathbb{P}\rightarrow 2^{\mathbb{Q}},\eta(p)=\{x\mid
\langle p,x\rangle \in \theta\}}$, derives all local variables of a
procedure. Let $\mathrm{\mathbb{N}=\{n_i\mid i\in N\}}$ be a pool of
control points from which we get unique control point during
pushdown rule derivation. In order to derive pushdown rules for
program, we define $\mathrm{\Phi(S,n_i,n_j,p,R)}$ in
Fig.~\ref{fig:derivation}. $n_i$ and $n_j$ are respectively the
entry and exit control point of $S$. $\rho=dom(\mu)$. $R$ is a set
collecting information about variable variation and variable
evaluation in statement transition for each pushdown rule. We extend
$R$ in \texttt{P-BRANCH} and \texttt{P-LOOP} to record the
precondition of control flow. $R$ is initialized with $\emptyset$.
$\varepsilon$ denotes the empty stack symbol.

In order to translate the derived pushdown rules into symbolic form,
we suppose the control point as explicit stack symbol. We prime
left-side variables of evaluation relations in $R$. Then we prime
the post-transition global variables and local variables in top
stack symbol, also double-prime the post-transition local variables
in bottom stack symbol. For the variable holding its value during
transition, extend $R$ with an equivalence relation of that
variable. Then we derive the symbolic form $\{\langle n_s\rangle
\hookrightarrow \langle (n_{t_1})\ldots (n_{t_k})\rangle \mid R\},
(k=0,1,2)$. Fig.~\ref{fig:example1} gives the symbolic pushdown
rules of the example calling procedure \texttt{func} in
Section~\ref{sec:intro}.

The derived symbolic pushdown system is sound on enforcing
noninterference for the core language programs as shown in the
following theorem. Let $\upmu=\langle \rho,\theta\rangle$ be the
state of $\mathcal {P}$. $\llbracket \mathcal {P}\rrbracket \upmu$
means an execution of $\mathcal {P}$ with initial state $\upmu$. It
returns a final state $\upmu'$ or $\bot$ if it does not terminate.

\begin{theorem}[Soundness\label{theorem:soundness}] Suppose
$\mathcal{P}=(P,\Gamma,\Delta)$ is the pushdown system of statement
$S$. $\Delta$ is derived by $\Phi(S,n_i,n_j,p_k,\emptyset)$.
$\forall \rho_0,\theta_0,\mu_0,\lambda_0,\llbracket \mathcal
{P}\rrbracket(\langle \rho_0,\theta_0\rangle )=\langle
\rho',\theta'\rangle$,
$(\mu_0,\lambda_0,S)\downarrow(\mu',\lambda')$. We have procedure
$p_k$ on the top of procedure stack of $\lambda_0$ and $\lambda'$.
Suppose $\forall l\in dom(\mu)$, $\mu_0(l)=\rho_0[l]$, and $\forall
l\in
dom(\lambda_0)$ corresponds to variable $x$ of $p_k$, $\lambda_0(l)=\eta_0(p_k)[x]$, we have\\
$\forall l\in dom(\mu)$, $\mu'(l)=\rho'[l]$, and $\forall l\in
dom(\lambda')$, if $l$ corresponds to variable $x$ of
$p_k$,$\lambda'(l)=\eta'(p_k)[x]$.
\end{theorem}
Proof: (See Appendix for details).

\section{\label{sec:prop_spec}Noninterference Property Specification}

In order to specify noninterference, we suppose the adversary can
observe whether program terminates. Then noninterference can be
classified into termination-sensitive(TS) and
termination-insensitive(TI). TS requires the correlative pairing
executions of program both terminate or both unterminate. TI only
judges the \textsf{low}-equivalence of final states when both
executions terminate, and does not get violated if either execution
unterminates. Let $L\subseteq \upmu$ be the public variables of
program. We define low-equivalent relation $=_L$ as $\forall x\in
\rho$, if $x\in L$, $\rho_1[x]=\rho_2[x]$, and $\forall \langle
p,y\rangle \in \theta$, if $\langle p,y\rangle \in L$,
$\eta_1(p)[y]=\eta_2(p)[y]$. Then TS and TI are formally defined as
follow.
\begin{definition}[TS]
A pushdown system $\mathcal {P}$ has a
property of \emph{TS} if it satisfies\\
\indent $\forall \upmu_1,\upmu_2$, $\upmu_1 =_L \upmu_2\wedge
\llbracket \mathcal {P}\rrbracket \upmu_1=\upmu_1' \Rightarrow
\exists \upmu_2', \llbracket \mathcal {P}\rrbracket
\upmu_2=\upmu_2'\wedge \upmu_1' =_L \upmu_2'$.
\end{definition}
\begin{definition}[TI]
A pushdown system $\mathcal {P}$ has a
property of \emph{TI} if it satisfies\\
$\forall \upmu_1,\upmu_2$, $\upmu_1 =_L \upmu_2\wedge \llbracket
\mathcal {P}\rrbracket \upmu_1=\upmu_1' \Rightarrow \llbracket
\mathcal {P}\rrbracket \upmu_2=\bot \vee (\exists \upmu_2',
\llbracket \mathcal {P}\rrbracket \upmu_2=\upmu_2'\wedge \upmu_1'
=_L \upmu_2')$.
\end{definition}
\indent TI can be simplified by restricting the precondition:
\begin{definition}[TI$'$]
A pushdown system $\mathcal {P}$ has a
property of \emph{TI} if it satisfies\\
\indent $\forall \upmu_1,\upmu_2$, $\upmu_1 =_L \upmu_2\wedge
\llbracket \mathcal {P}\rrbracket \upmu_1=\upmu_1'\wedge \llbracket
\mathcal {P}\rrbracket \upmu_2=\upmu_2' \Rightarrow \upmu_1' =_L
\upmu_2'$.
\end{definition}
For simplicity, we only consider TI in the following. The LTL
property w.r.t TS can be derived similarly by method in
\cite{DBLP:conf/csfw/BartheDR04}.

\subsection{\label{subsec:osc}Ordinary Self-Composition}

The primary motivation of self-composition is to model two
correlative runs of program with indistinguishable inputs as a
single run of composed program, which greatly benefits the
algorithmic verification techniques, such as model checking, on the
property expressing via temporal logics.

Suppose $\mathcal {P}_1$ and $\mathcal {P}_2$ are pushdown systems.
$\upmu_1$, $\upmu_1'$ are states of $\mathcal {P}_1$, and $\upmu_2$,
$\upmu_2'$ are states of $\mathcal {P}_2$. Let $\upmu_1 \cap
\upmu_2=(\rho_1 \cap \rho_2)\cup (\theta_1 \cap \theta_2)$. $\oplus$
is disjoint union of memory. If $\upmu_1\cap \upmu_2=\emptyset$, we
define \emph{disjoint composition} operation $\vartriangleright$ of
$\mathcal {P}_1$ and
$\mathcal {P}_2$ as\\
$\llbracket \mathcal {P}_1\rrbracket(\upmu_1\oplus
\upmu)=(\upmu_1'\oplus \upmu)$ for some $\upmu$, and $\llbracket
\mathcal {P}_2\rrbracket(\upmu'\oplus \upmu_2)=(\upmu'\oplus
\upmu_2')$ for some $\upmu'$, iff $\llbracket \mathcal
{P}_1\vartriangleright \mathcal {P}_2\rrbracket (\upmu_1\oplus
\upmu_2)=(\upmu_1'\oplus \upmu_2')$.

Suppose $\upmu=\langle \rho,\theta\rangle $ be the state of
$\mathcal {P}$. The state of $\mathcal{P}[\xi]$ is derived by
renaming the variables of $\mathcal{P}$ by function
$\xi:\upmu\rightarrow \langle \{x^* \mid x\in \rho\},\{\langle
p,y^*\rangle \mid \langle p,y\rangle \in \theta\}\rangle $, such
that $\forall x\in \rho, \xi(x)=x^*$ and $\forall \langle p,y\rangle
\in \theta, \xi(\langle p,y\rangle )=\langle p,y^*\rangle $. The
ordinary self-composition is defined based on disjoint composition
and variable renaming, and the corresponding TI is defined as:
\begin{definition}[TI, ordinary self-composition]
$\mathcal {P}$ has a property of TI iff\\
\hspace*{\fill}$\upmu_1 =_L \upmu_2 \wedge \llbracket
\mathcal{P}\vartriangleright \mathcal{P}[\xi]\rrbracket(\upmu_1
\oplus \upmu_2)=(\upmu_1' \oplus
\upmu_2')\Rightarrow  \upmu_1' =_L \upmu_2'$,\hspace*{\fill}\\
where $\upmu_1,\upmu_1'$ are states of $\mathcal {P}$, and
$\upmu_2,\upmu_2'$ are states of $\mathcal {P}[\xi]$.
\end{definition}

The corresponding self-composing algorithm on pushdown system to
derive $\mathcal{P}\vartriangleright \mathcal{P}[\xi]$ is proposed
as follow.
\begin{compactenum}[1.]
\item Derive $\mathcal {P}[\xi]$ by substituting variables of $\mathcal {P}$ with the corresponding renamed
variables defined by $\xi$, and substituting each control point
$n_i$ of $\mathcal {P}$ with $n_i^*$.
\item Merge the pushdown rules of $\mathcal {P}$ and $\mathcal {P}[\xi]$.
\item Modify the last pushdown rule of $\mathcal {P}$ from
$\langle n_{final}\rangle \hookrightarrow \langle
\varepsilon\rangle$ to $\langle n_{final}\rangle \hookrightarrow
\langle n_{init}^*\rangle$, and for the totalness of composed
pushdown system, modify the last pushdown rule of $\mathcal{P}[\xi]$
from $\langle n_{final}^*\rangle \hookrightarrow
\langle\varepsilon\rangle$ to $\langle
n_{final}^*\rangle\hookrightarrow\langle n_{final}^*\rangle$.
Algorithm~\ref{algo:last_trans} is used to find the last transition
of original pushdown system.
\item For each rule of $\mathcal{P}$, $\forall x^*\in \rho_2$, add
${x^*}'=x^*$ to $R$. For each $\langle n_i\rangle \hookrightarrow
\langle n_j\rangle$ of $\mathcal{P}$, $\forall \langle p,y^*\rangle
\in \theta_2$, add $\eta(p)[y^*]'=\eta(p)[y^*]$ to $R$. For each
$\langle n_i\rangle \hookrightarrow \langle n_k,n_j\rangle$ of
$\mathcal{P}$, $\forall \langle p,y^*\rangle \in \theta_2$, add
$\eta(p)[y^*]''=\eta(p)[y^*]$ to $R$.
\item For each rule of $\mathcal{P}[\xi]$, $\forall x\in \rho_1$, add
$x'=x$ to $R$. For each $\langle n_i^*\rangle \hookrightarrow
\langle n_j^*\rangle$ of $\mathcal{P}[\xi]$, $\forall \langle
p,y\rangle \in \theta_1$, add $\eta(p)[y]'=\eta(p)[y]$ to $R$. For
each $\langle n_i^*\rangle \hookrightarrow \langle
n_k^*,n_j^*\rangle$ of $\mathcal{P}[\xi]$, $\forall \langle
p,y\rangle \in \theta_1$, add $\eta(p)[y]''=\eta(p)[y]$ to $R$.
\item $\forall \langle
p,y\rangle \in \theta_1$, add $\eta(p)[y]'=\eta(p)[y]$ to $R$ of
$\langle n_{final}\rangle \hookrightarrow \langle
n_{init}^*\rangle$.
\end{compactenum}
Then TI can be expressed by linear temporal logic and verified on
this composed pushdown system by model checking. $\mathcal {P}$ has
a property of TI if and only if $\mathcal {P}\vartriangleright
\mathcal {P}[\xi]$ satisfies $(\upmu_1 =_L \upmu_2)\Rightarrow
\textsf{G}(n_{final}^*\Rightarrow (\upmu_1 =_L \upmu_2))$.

\begin{algorithm}
\caption{\label{algo:last_trans}LastTransitionFinding}
$toVisit\mathrel{\mathop:}=startConf\langle p,\gamma_0\rangle.\gamma_0$; $visited\mathrel{\mathop:}=\emptyset$;\\
\While {$toVisit\neq \emptyset$} {
    $cur\mathrel{\mathop:}=toVisit.head$; $toVisit\mathrel{\mathop:}=toVisit\setminus\{cur\}$;\\
    \ForAll {$t\in trans \neq \emptyset$} {
        \uIf {$t.expr=\langle p,cur\rangle\hookrightarrow \langle p, \gamma', \gamma''\rangle\ \wedge \neg find(\gamma'',visited)$} {$toVisit\mathrel{\mathop:}=toVisit\cup\{\gamma''\}$;}
        \uIf {$t.expr=\langle p,cur\rangle \hookrightarrow \langle p,\gamma'\rangle \wedge \neg find(\gamma',visited)$} {$toVisit\mathrel{\mathop:}=toVisit\cup\{\gamma'\}$;}
        \uIf {$t.expr=\langle p,cur\rangle \hookrightarrow \langle p,\epsilon \rangle$} {\Return $t$;}
    }
    $visited\mathrel{\mathop:}=visited\cup\{cur\}$;
}
\end{algorithm}

\subsection{\label{subsec:csc}Compact Self-Composition}

Since our approach is based on symbolic model checking, the variable
count has great impact on the size of \emph{binary decision diagram}
(BDD) and performance of model checker. The increase in variable
count after ordinary self-composition seriously decreases the
efficiency of model checking. This motivates us to find new
self-composing methods in inter-procedural settings. The compact
form of self-composition relies on the following two assumptions
\begin{compactitem}
\item All the variables observable by the adversary are global. That
means to treat the procedure with observation point in it as
\emph{main} procedure.
\item All local variables are initialized before being used
in procedure, and vanish while procedure returns, thus are
considered as \textsf{high}.
\end{compactitem}
Let
$\upmu_1\oplus_l\upmu_2=\upmu_1\oplus\rho_2=\langle\rho_1,\theta_1\rangle\oplus\rho_2$,
and $\upmu_1\oplus_r\upmu_2=\rho_1\oplus\upmu_2=\rho_1\oplus\langle
\rho_2,\theta_2\rangle$. Suppose $\oplus_\rho$ is the disjoint union
of global variables, we have \emph{compact disjoint composition}
$\vartriangleright_\rho$ of $\mathcal{P}_1$ and $\mathcal{P}_2$,
where $\llbracket \mathcal {P}_1\rrbracket(\upmu_1\oplus_l
\upmu)=(\upmu_1'\oplus \upmu)$ for some $\upmu$, and $\llbracket
\mathcal {P}_2\rrbracket(\upmu'\oplus_r \upmu_2)=(\upmu'\oplus
\upmu_2')$ for some $\upmu'$, iff $\llbracket \mathcal
{P}_1\vartriangleright_\rho \mathcal {P}_2\rrbracket
(\upmu_1\oplus_\rho \upmu_2)=(\upmu_1'\oplus_\rho \upmu_2')$.
\begin{definition}[TI,compact self-composition]
$\mathcal {P}$ has a property of TI iff\\
\hspace*{\fill}$\rho_1 =_L \rho_2 \wedge \llbracket
\mathcal{P}\vartriangleright_\rho \mathcal{P}[\xi]\rrbracket(\upmu_1
\oplus_\rho \upmu_2)=(\upmu_1' \oplus_\rho \upmu_2')\Rightarrow
\rho_1' =_L \rho_2'$\hspace*{\fill}
\end{definition}

The self-composing algorithm w.r.t compact form is derived by
modifying step 4 to step 6 of the algorithm in Section
\ref{subsec:osc} to the following strategies
\begin{compactenum}
\item[4$'$.] For each rule of $\mathcal{P}$, $\forall x^*\in \rho_2$, add
${x^*}'=x^*$ to $R$.
\item[5$'$.] For each rule of $\mathcal{P}[\xi]$, $\forall x\in \rho_1$, add
$x'=x$ to $R$.
\end{compactenum}

A unified algorithm of ordinary and compact forms of
self-composition is illustrated by Algorithm~\ref{algo:sc}. Suppose
$\mathcal {P}.lvars:CP\mapsto 2^{\mathbb{Q}}$ be a mapping from the
set of control points to a set of local variables related to certain
procedure. Also suppose procedure tag $p_i$ is represented by the
set of control points arrived during execution of this procedure.
Then we have $\mathcal{P}.lvars(\gamma)=\{x|\gamma \in p_i,\langle
p_i, x\rangle\in\theta\}$.

\begin{algorithm}
\caption{\label{algo:sc} Ordinary and Compact Self Composition}

\KwData{$\mathcal{P},sc\_mode$}
\KwResult{$\mathcal{P}\vartriangleright\mathcal{P}[\xi]$}

\Begin{

\ForAll {$x\in \mathcal{P}.gvars$} {
    $\mathcal{P}[\xi].gvars\mathrel{\mathop:}=\mathcal{P}[\xi].gvars\cup \{\xi(x)\}$;
}

$(\mathcal{P}\vartriangleright
\mathcal{P}[\xi]).gvars\mathrel{\mathop:}=\mathcal{P}.gvars\cup
\mathcal{P}[\xi].gvars$;

\ForAll {$\gamma\in CP$} {
    \lForAll {$x\in \mathcal{P}.lvars(\gamma)$} {
            $\mathcal{P}[\xi].lvars(\gamma t)\mathrel{\mathop:}=\mathcal{P}[\xi].lvars(\gamma t)\cup \{\xi(x)\}$;
    }
}

$\mathcal{P}[\xi].startConf\mathrel{\mathop:}=\langle
\mathcal{P}.startConf.p, Append(\mathcal{P}.startConf.\gamma,\
't')\rangle$;
$(\mathcal{P}\vartriangleright\mathcal{P}[\xi]).startConf\mathrel{\mathop:}=\mathcal{P}.startConf$;

\ForAll {$t\in \mathcal{P}.trans$} {
    \uIf {$t.expr=\langle p,\gamma\rangle\hookrightarrow \langle p, \gamma' \gamma''\rangle$} {
        \uIf {$sc\_mode=ordinary$} {
            $\mathcal{P}[\xi].trans\mathrel{\mathop:}=\mathcal{P}[\xi].trans\cup\{\langle p,\gamma t\rangle\hookrightarrow\langle
            p, \gamma't, \gamma''t\rangle|$\\
            \hspace*{\fill} $(t.rel){\xi(x)\atop x}\wedge\bigwedge_{x_i\in\mathcal{P}.gvars}(x_i'=x_i)\wedge\bigwedge_{x_i\in\mathcal{P}.lvars(\gamma)}(x_i''=x_i)
            \}$;
            $t.rel\mathrel{\mathop:}=t.rel\wedge\bigwedge_{x_i\in\mathcal{P}[\xi].gvars}(x_i'=x_i)\wedge\bigwedge_{x_i\in\mathcal{P}[\xi].lvars(\gamma
            t)}(x_i''=x_i)$;
        }
        \uIf {$sc\_mode=compact$} {
            $\mathcal{P}[\xi].trans\mathrel{\mathop:}=\mathcal{P}[\xi].trans\cup\{\langle p,\gamma t\rangle\hookrightarrow\langle
            p, \gamma't, \gamma''t\rangle|$\\
            \hspace*{2cm}$(t.rel){\xi(x)\atop x}\wedge\bigwedge_{x_i\in\mathcal{P}.gvars}(x_i'=x_i)\}$;
            $t.rel\mathrel{\mathop:}=t.rel\wedge\bigwedge_{x_i\in\mathcal{P}[\xi].gvars}(x_i'=x_i)$;
        }
    }
    \uElseIf {$t.expr=\langle p,\gamma\rangle\hookrightarrow \langle p, \gamma'\rangle$} {
        \uIf {$sc\_mode=ordinary$} {
            $\mathcal{P}[\xi].trans\mathrel{\mathop:}=\mathcal{P}[\xi].trans\cup\{\langle p,\gamma t\rangle\hookrightarrow\langle
            p, \gamma't\rangle|$\\
            \hspace*{2cm}$(t.rel){\xi(x)\atop x}\wedge\bigwedge_{x_i\in\mathcal{P}.gvars\cup\mathcal{P}.lvars(\gamma)}(x_i'=x_i)\}$;
            $t.rel\mathrel{\mathop:}=t.rel\wedge\bigwedge_{x_i\in\mathcal{P}[\xi].gvars\cup \mathcal{P}[\xi].lvars(\gamma t)}(x_i'=x_i)$;
        }
        \uIf {$sc\_mode=compact$} {
            $\mathcal{P}[\xi].trans\mathrel{\mathop:}=\mathcal{P}[\xi].trans\cup\{\langle p,\gamma t\rangle\hookrightarrow\langle
            p, \gamma't\rangle|$\\
            \hspace*{2cm}$(t.rel){\xi(x)\atop x}\wedge\bigwedge_{x_i\in\mathcal{P}.gvars}(x_i'=x_i)\}$;
            $t.rel\mathrel{\mathop:}=t.rel\wedge\bigwedge_{x_i\in\mathcal{P}[\xi].gvars}(x_i'=x_i)$;
        }
    }
    \Else(\tcc*[f]{$t.expr=\langle p,\gamma\rangle\hookrightarrow \langle p,\epsilon\rangle$}){
        \uIf {$t=LastTransitionFinding(\mathcal{P})$} {
            $\mathcal{P}[\xi].trans\mathrel{\mathop:}=\mathcal{P}[\xi].trans\cup\{\langle p,\gamma t\rangle\hookrightarrow\langle p,\gamma t\rangle
            |$\\
            \hspace*{2cm}$(t.rel){\xi(x)\atop x}\wedge\bigwedge_{x_i\in\mathcal{P}.gvars}(x_i'=x_i)\}$;\\
            \uIf {$sc\_mode=ordinary$} {
                $\mathcal{P}.trans\mathrel{\mathop:}=(\mathcal{P}.trans\setminus\{t\})\cup\{\langle
                p,\gamma\rangle\hookrightarrow\mathcal{P}[\xi].startConf|$\\
                \hspace*{\fill}$(t.rel)\wedge\bigwedge_{x_i\in\mathcal{P}.lvars(\gamma)\cup\mathcal{P}[\xi].gvars\cup\mathcal{P}[\xi].lvars(\gamma t)}(x_i'=x_i)\}$;
            }
            \uIf {$sc\_mode=compact$} {
                $\mathcal{P}.trans\mathrel{\mathop:}=(\mathcal{P}.trans\setminus\{t\})\cup\{\langle
                p,\gamma\rangle\hookrightarrow\mathcal{P}[\xi].startConf|$\\
                \hspace*{1.6cm}$(t.rel)\wedge\bigwedge_{x_i\in\mathcal{P}[\xi].gvars}(x_i'=x_i)\}$;
            }
        }
        \Else(\tcc*[f]{last transition of other procedures except main}){
            $\mathcal{P}[\xi].trans\mathrel{\mathop:}=\mathcal{P}[\xi].trans\cup\{\langle p,\gamma t\rangle\hookrightarrow\langle p,
            \epsilon\rangle|$\\
            \hspace*{2cm}$(t.rel){\xi(x)\atop x}\wedge\bigwedge_{x_i\in\mathcal{P}.gvars}(x_i'=x_i)\}$;
            $t.rel\mathrel{\mathop:}=t.rel\wedge\bigwedge_{x_i\in\mathcal{P}[\xi].gvars}(x_i'=x_i)$;
        }
    }
}

$(\mathcal{P}\vartriangleright\mathcal{P}[\xi]).trans\mathrel{\mathop:}=\mathcal{P}.trans\cup\mathcal{P}[\xi].trans$;\\
\uIf {$sc\_mode=ordinary$} {
    \ForAll {$x\in\mathcal{P}.lvars(\gamma)$} {
        $\mathcal{P}.lvars(\gamma)\mathrel{\mathop:}=\mathcal{P}.lvars(\gamma)\cup\{\xi(x)\}$;
        $\mathcal{P}[\xi].lvars(\gamma t)\mathrel{\mathop:}=\mathcal{P}[\xi].lvars(\gamma t)\cup\{x\}$;
    }
}

\ForAll {$\gamma\in CP$} {
    $(\mathcal{P}\vartriangleright\mathcal{P}[\xi]).lvars\mathrel{\mathop:}=(\mathcal{P}\vartriangleright\mathcal{P}[\xi]).lvars\cup\{\gamma\mapsto
    \mathcal{P}.lvars(\gamma), \gamma
    t\mapsto\mathcal{P}[\xi].lvars(\gamma t)\}$;

}

}
\end{algorithm}

\subsection{\label{subsec:csc2}Contracted Self-Composition}

According to the assumptions given in Section \ref{subsec:csc}, we
know that noninterference property at certain observation point of
program does not really care the value of local variables in the
callee procedures. By the algorithm of Section \ref{subsec:csc}, we
have reduced the length of $R$ accompanying pushdown rules. But what
if we want to reduce the states of composed program? Since the
adversary can only observe the global variables of \emph{main}
procedure, we can avoid duplicating memory of local variables by
making the composed part of \emph{main} procedure call the original
callee, instead of the composed callee, as original part of
\emph{main} procedure does. This form of self-composition
additionally relies on the following assumption:
\begin{compactitem}
\item Global variable can not be used in callee procedure unless
it is passed as parameter of callee procedure.
\end{compactitem}
With the contracted form of self-composition, we do not compose
callee procedures. Therefore we could not express the direct effect
of newly defined global variables on local variables of callee
procedures. The state of $\mathcal{P}[\xi']$ is derived by
$\xi':\langle\rho,\theta\rangle\rightarrow\langle\{x^* \mid x\in
\rho\},\theta\rangle$ such that $\forall x\in \rho, \xi'(x)=x^*$ and
$\forall \langle p,y\rangle \in \theta, \xi'(\langle p,y\rangle
)=\langle p,y\rangle $. We have TI w.r.t a contracted form.
\begin{definition}[TI,contracted self-composition]
$\mathcal {P}$ has a property of
TI iff\\
\hspace*{\fill}$\rho_1 =_L \rho_2 \wedge \llbracket
\mathcal{P}\vartriangleright_\rho
\mathcal{P}[\xi']\rrbracket(\upmu_1 \oplus_\rho \upmu_2)=(\upmu_1'
\oplus_\rho \upmu_2')\Rightarrow \rho_1' =_L \rho_2'$\hspace*{\fill}
\end{definition}
\indent The related self-composing algorithm distinguishes pushdown
rules of \emph{main} procedure from pushdown rules of callee
procedures.
\begin{compactenum}
\item Derive $\mathcal{P}[\xi']$ by substituting variables of
$\mathcal{P}$ with renamed variables defined by $\xi'$. For pushdown
rules of \emph{main} procedure, substitute each $\langle n_i\rangle
\hookrightarrow \langle n_j\rangle$ with $\langle n_i^*\rangle
\hookrightarrow \langle n_j^*\rangle$, $\langle n_i\rangle
\hookrightarrow \langle \varepsilon\rangle$ with $\langle
n_i^*\rangle \hookrightarrow \langle \varepsilon\rangle$, and
$\langle n_i\rangle \hookrightarrow \langle n_k,n_j\rangle$ with
$\langle n_i^*\rangle \hookrightarrow \langle n_k,n_j^*\rangle$.
\item Merge $\mathcal{P}[\xi']$ and $\mathcal{P}$, taking duplicated
pushdown rules only once.
\item Same as step 3 in Section \ref{subsec:osc}.
\item Same as step 4$'$ and 5$'$ in Section \ref{subsec:csc}.
\end{compactenum}
\indent The corresponding LTL-expressed TI for compact
self-composition and contracted self-composition is $(\rho_1 =_L
\rho_2)\Rightarrow \textsf{G}(n_{final}^*\Rightarrow (\rho_1 =_L
\rho_2))$. Let $RT$ be the set of global variables used as the store
of return value of the callee procedures. The algorithm is described
in detail by Algorithm~\ref{algo:sc_con}. Our experiments in Section
\ref{sec:expr} illustrate the improvement on efficiency brought by
these derived forms of self-composition.

\begin{algorithm}
\caption{\label{algo:sc_con} Contracted Self Composition}

\KwData{$\mathcal{P}$}
\KwResult{$\mathcal{P}\vartriangleright_\rho\mathcal{P}[\xi']$}

\Begin{

\ForAll {$x\in \mathcal{P}.gvars\setminus RT$} {
    $\mathcal{P}[\xi'].gvars\mathrel{\mathop:}=\mathcal{P}[\xi'].gvars\cup \{\xi'(x)\}$;
}

$(\mathcal{P}\vartriangleright_\rho
\mathcal{P}[\xi']).gvars\mathrel{\mathop:}=\mathcal{P}.gvars\cup
\mathcal{P}[\xi'].gvars$;

$(\mathcal{P}\vartriangleright_\rho\mathcal{P}[\xi']).lvars\mathrel{\mathop:}=\mathcal{P}.lvars$;

\ForAll {$\gamma\in CP_{Main}$} {
    \lForAll {$x\in \mathcal{P}.lvars(\gamma)$}
    {$(\mathcal{P}\vartriangleright_\rho\mathcal{P}[\xi']).lvars\mathrel{\mathop:}=(\mathcal{P}\vartriangleright_\rho\mathcal{P}[\xi']).lvars\cup\{\gamma t\mapsto x\}$;}
}

$\mathcal{P}[\xi'].startConf\mathrel{\mathop:}=\langle
\mathcal{P}.startConf.p, Append(\mathcal{P}.startConf.\gamma,\
't')\rangle$;
$(\mathcal{P}\vartriangleright_\rho\mathcal{P}[\xi']).startConf\mathrel{\mathop:}=\mathcal{P}.startConf$;

\ForAll {$t\in MainTrans(\mathcal{P}.trans)$} {
    \uIf {$t.expr=\langle p,\gamma\rangle\hookrightarrow\langle p,\epsilon\rangle$} {
        $\mathcal{P}[\xi'].trans\mathrel{\mathop:}=\mathcal{P}[\xi'].trans\cup\{\langle p,\gamma t\rangle\hookrightarrow\langle p,\gamma t\rangle|$
        \hspace*{2cm}$(t.rel){\xi'(x)\atop x\in\mathcal{P}.gvars\setminus RT}\wedge\bigwedge_{x_i\in\mathcal{P}.gvars\setminus RT}(x_i'=x_i)\}$;
    }
    \uElseIf {$t.expr=\langle p,\gamma\rangle\hookrightarrow\langle p,\gamma'\rangle$} {
        $\mathcal{P}[\xi'].trans\mathrel{\mathop:}=\mathcal{P}[\xi'].trans\cup\{\langle p,\gamma t\rangle\hookrightarrow\langle p,\gamma't\rangle|$
        \hspace*{2cm}$(t.rel){\xi'(x)\atop x\in\mathcal{P}.gvars\setminus RT}\wedge\bigwedge_{x_i\in\mathcal{P}.gvars\setminus RT}(x_i'=x_i)\}$;
    }
    \Else (\tcc*[f]{$t.expr=\langle p,\gamma\rangle\hookrightarrow\langle p,\gamma',\gamma''\rangle$}) {
        $\mathcal{P}[\xi'].trans\mathrel{\mathop:}=\mathcal{P}[\xi'].trans\cup\{\langle p,\gamma t\rangle\hookrightarrow\langle p,\gamma',\gamma''t\rangle|$
        \hspace*{2cm}$(t.rel){\xi'(x)\atop x\in\mathcal{P}.gvars\setminus RT}\wedge\bigwedge_{x_i\in\mathcal{P}.gvars\setminus RT}(x_i'=x_i)\}$;
    }
    $t.rel\mathrel{\mathop:}=t.rel\wedge\bigwedge_{x_i\in\mathcal{P}[\xi'].gvars}(x_i'=x_i)$;
}

$(\mathcal{P}\vartriangleright_\rho\mathcal{P}[\xi']).trans\mathrel{\mathop:}=\mathcal{P}.trans\cup\mathcal{P}[\xi'].trans$;

}
\end{algorithm}

\begin{algorithm}
\caption{\label{algo:main_trans}MainTrans (used in
Algorithm~\ref{algo:sc_con})}

$toVisit\mathrel{\mathop:}=startConf\langle p,\gamma_0\rangle.\gamma_0$; $visited\mathrel{\mathop:}=\emptyset$; $r\_trans\mathrel{\mathop:}=\emptyset$;\\
\While {$toVisit\neq \emptyset$} {
    $cur\mathrel{\mathop:}=toVisit.head$; $toVisit\mathrel{\mathop:}=toVisit\setminus\{cur\}$;\\
    \ForAll {$t\in trans \neq \emptyset$} {
        \uIf {$t.expr=\langle p,cur\rangle\hookrightarrow \langle p, \gamma', \gamma''\rangle\ \wedge \neg find(\gamma'',visited)$} {
            $toVisit\mathrel{\mathop:}=toVisit\cup\{\gamma''\}$; $r\_trans\mathrel{\mathop:}=r\_trans\cup t$;
        }
        \uIf {$t.expr=\langle p,cur\rangle \hookrightarrow \langle p,\gamma'\rangle \wedge \neg find(\gamma',visited)$} {
            $toVisit\mathrel{\mathop:}=toVisit\cup\{\gamma'\}$; $r\_trans\mathrel{\mathop:}=r\_trans\cup t$;
        }
        \uIf {$t.expr=\langle p,cur\rangle \hookrightarrow \langle p,\epsilon \rangle$} {
            $r\_trans\mathrel{\mathop:}=r\_trans\cup t$;
        }
    }
    $visited\mathrel{\mathop:}=visited\cup\{cur\}$;
}

\Return {$r\_trans$;}
\end{algorithm}

\section{\label{sec:expr}Experiments}

The main purpose of our experiments is to reveal that the derived
forms of self-composition, compared with the ordinary form, can
really improve the efficiency of verification. Also we make clear
how we could benefit from the procedural settings and model
checking.

We have implemented all three forms of self-composition for symbolic
pushdown system. This is a static transformation before the pushdown
system parsing phase. Self-composed pushdown system is generated and
related Moped options are set. We add command-line options to Moped
version 1. We also implicitly require the derived variables and
control points in composed program tagged by postfix \emph{-t},
especially the final control point named as \emph{\_final}. With
this assumption, user can express TI/TS noninterference by LTL from
the original symbolic pushdown system instead of the one after
self-composition. All experiments were performed on a laptop with
1.66 GHz Intel Core 2 CPU and 1 GB RAM running Linux Fedora Core 6.
The BDD library used by Moped is CUDD 2.3.1. Experimental results
are partially presented in Table~\ref{tab:expr}.

\begin{figure}[t]
\includegraphics
[scale=.7,bb=55 285 560 520]{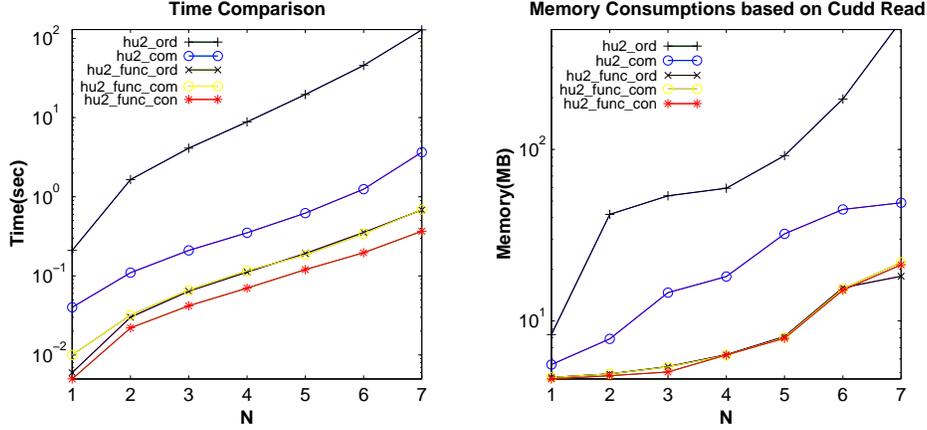}
\caption{\label{fig:hu_result}Time and Memory Comparison}
\end{figure}

\begin{figure}[t]
\hspace*{1cm}
\includegraphics
[scale=.6,bb=0 0 460 330]{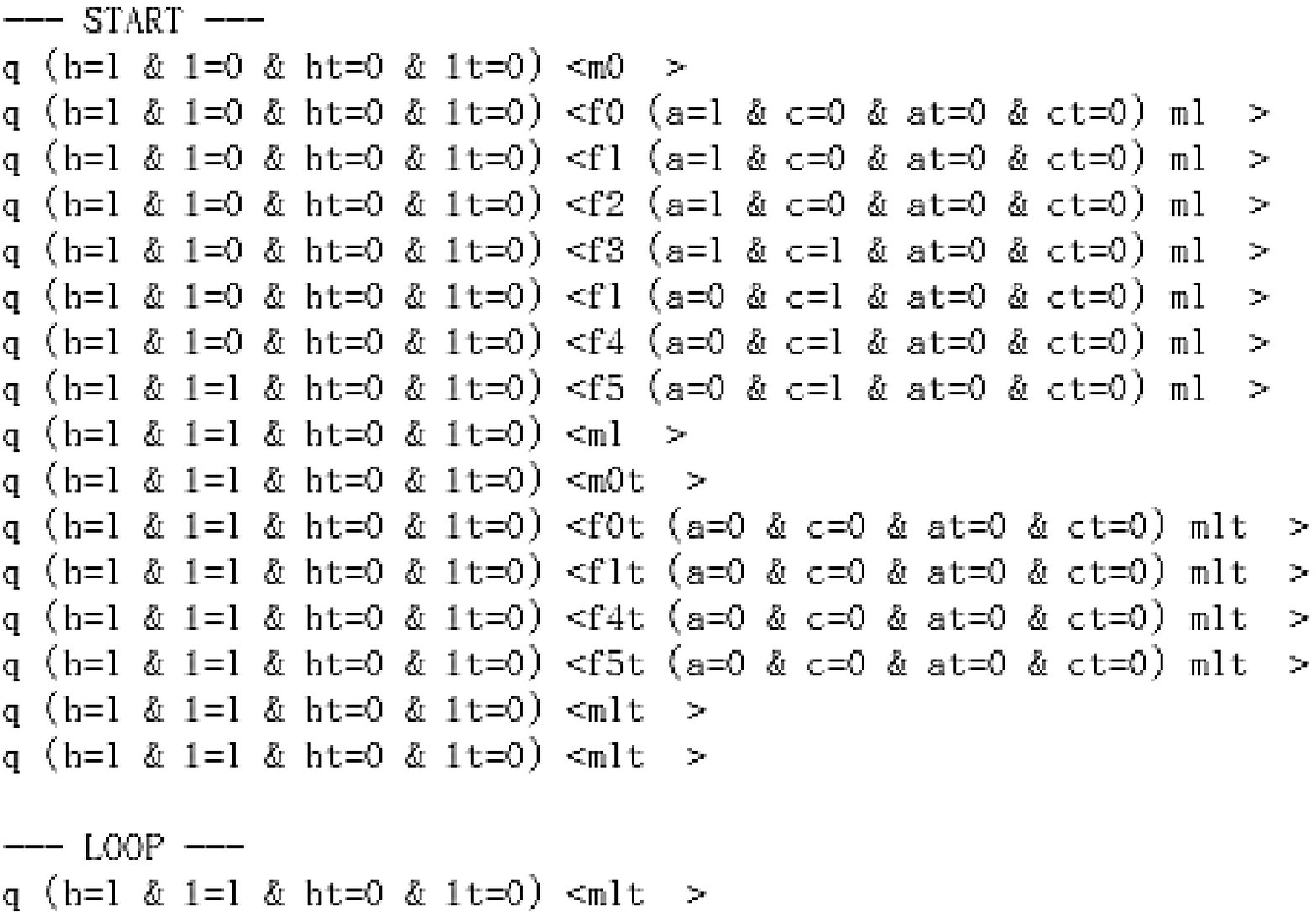}
\caption{\label{fig:snapshot}Snapshot of Counterexample}
\end{figure}

In this table, \#gvars and \#lvars give the number of global
variables and the number of local variables respectively. \#pubs
represents the number of public/\textsf{low} variables. $N$ gives
the number of bit of each integer variable. As shown in
Fig.~\ref{fig:hu_result}, larger $N$ causes increase in consumed
time and memory. Fortunately however, the secure/insecure judgement
made by our method is mostly insensitive to $N$, thus we could set
$N$ as small as possible to outperform other methods. The only
matter is to ensure the range of integer ($0\sim 2^N-1$) cover the
constant value assigned to it. The basic checking result
\texttt{Yes} reports the program is secure, while the answer
\texttt{No}, along with the witness path generated as a
counterexample, not only tells the program is insecure, but
facilitates our method on reasoning the flow path from certain
\textsf{high} variable to \textsf{low} variables. In another word,
the single counterexample identified by model checker could tell us
the flow-source variable with security level \textsf{high}. This is
the first step to fix the flaw of program, and probably the most
obvious benefit provided by model checking. Consider again the
program calling \textsf{func} in Section \ref{sec:intro}, the
snapshot is presented in Fig.~\ref{fig:snapshot} where $N=1$. We can
clearly observe the difference of $l$ and $lt$ at observation point
comes from the difference on value of $h$ and $ht$ at the beginning
of execution. The illegal flow is performed by transition rule from
\texttt{f4} to \texttt{f5}.

\begin{table*}[t]
\renewcommand{\arraystretch}{1.1}
\begin{tabular}{|l||c|c|c|c|c|c||c|c|c|c|}\hline
Prog & $\mid$Prog$\mid$ & \#gvars & \#lvars & \#pubs & Yes & N &
\multicolumn{2}{c|}{ordinary} &
\multicolumn{2}{c|}{compact}\\\cline{8-11}

& & & & & /No & & Time(s) & Mem(MB) & Time(s) &
Mem(MB)\\\hline\hline

ttaa1 & 13 & 3 & 2 & 2 & Yes & 9 & 0.01 & 45.82 & 0.01 &
45.58\\\hline

ttaa2 & 14 & 3 & 2 & 2 & No & 8 & 0.16 & 41.99 & 0.03 &
41.95\\\hline

ttaa3 & 22 & 4 & 3 & 3 & Yes & 8 & 1.86 & 49.61 & 1.52 &
48.87\\\cline{7-11}

& & & & & & 1 & $<$0.01 & 4.74 & $<$0.01 & 4.71\\\hline

lgj & 13 & 3 & 1 & 2 & Yes & 9 & $<$0.01 & 4.65 & $<$0.01 &
4.65\\\hline

func1 & 12 & 2 & 2 & 1 & No & 8 & 4.45 & 36.68 & 2.48 &
19.02\\\hline

hu1 & 54 & 2 & 33 & 1 & No & 1 & -- & -- & 12.10 & 73.19\\\hline

hu2 & 70 & 2 & 17 & 1 & No & 3 & 4.11 & 53.61 & 0.21 & 14.62\\\hline

hu2\_func & 26 & 3 & 1 & 1 & No & 8 & 1.44 & 35.63 & 1.40 &
34.92\\\hline

hu3 & 63 & 60 & 0 & 20 & No & 1 & --- & --- & 0.07 & 202.56\\\hline

hu3* & 63 & 60 & 0 & 40 & No & 1 & --- & --- & 128.16 &
1015.61\\\hline

mod\_imap & 526 & 9 & 13 & 8 & No & 4 & 14.00 & 34.55 & 11.25 &
32.88\\\cline{7-11}

& & & & & & 5 & 404.79 & 65.40 & 327.93 & 55.09\\\hline
\end{tabular}
\caption{\label{tab:expr}Experimental Results}
\end{table*}

Program \texttt{ttaa1}, \texttt{ttaa2} and \texttt{ttaa3} are
respectively programs presented in Figure~1, Figure~4 and Figure~9
of \cite{DBLP:conf/sas/TerauchiA05}. In
\cite{DBLP:conf/sas/TerauchiA05} the authors report a failure on
verifying \texttt{ttaa3} secure using BLAST. Here using our method
we can verify this program secure. \texttt{hu1}, \texttt{hu2} and
\texttt{hu3} model the program \texttt{expr1.c}, \texttt{expr2.c}
and \texttt{expr4.c} of \cite{DBLP:conf/pldi/UnnoKY06}. In order to
explain the benefit from procedural setting, we encapsulate the
\textrm{if}-branch of \texttt{hu2} into a procedure and change
\texttt{hu2} to a program calling this procedure for sixteen times.
The resulting program is \texttt{hu2\_func}. The evaluation results
are illustrated by Fig.~\ref{fig:hu_result}. Compared with ordinary
self-composition of \texttt{hu2}, consumed time and memory are
greatly reduced by the ordinary self-composition of
\texttt{hu2\_func}. To make clear the effectiveness of the derived
forms of self-composition, we did ordinary and compact
self-composition on \texttt{hu2} (since no procedure exists, the
contracted form is not available) and all three forms of
self-composition on \texttt{hu2\_func}. The time reduction by using
compact form is more notable on \texttt{hu2} than on
\texttt{hu2\_func}. This is because \texttt{hu2} has more local
variables, and the conjunction of relations in $R$ is much shorter
for compact form than ordinary form. That means each symbolic
pushdown rule of derived pushdown system corresponds to more
explicit pushdown rules and smaller BDD. The only one local variable
of \texttt{hu2\_func} could not make this effect obvious. The
contracted form could outperform the compact form because the number
of derived local variables is reduced to half compared with compact
self-composition and the number of pushdown rules is also reduced.
\texttt{hu3} can be verified insecure in two aspects. First, it can
leak path information of whether branch with condition
$\mathrm{b_{10}}$ is taken. Here we need to specify the initial
value of $\mathrm{low_i}(i=1,2,\ldots,20)$ to be 0 (See \texttt{hu3}
in Table~\ref{tab:expr}). Second, we can reveal that
$\mathrm{low_{10}}$ leaks $\mathrm{high_{10}}$ by specifying the
original part and the composed part of derived pushdown system take
the same path ($\mathrm{b_i=bt_i, i=1,2,\ldots,20}$) (See
\texttt{hu3*} in Table~\ref{tab:expr}). Here the path conditions
$\mathrm{b_i}$ should be global. Without any local variable, the
ordinary form and compact form of \texttt{hu3} are of no difference
and we record experiment results of either instead of both in
Table~\ref{tab:expr}.

To show the realistic usage of our method, we have verified the
standard module \texttt{mod\_imap.c} of Apache 1.3.23. This program
is also verified as an important experiment in
\cite{DBLP:conf/pldi/UnnoKY06}. In the \texttt{imap\_url} function
of this program, a possible tainted \texttt{Referer} url could be
returned and passed to \texttt{menu\_$\ast$} functions and then to
the arguments of \texttt{ap\_rvputs}. \texttt{ap\_rvputs} passes
these arguments to client browser and causes a \emph{cross-site
scripting} (XSS) vulnerability. To check cross-site scripting, we
need to consider the parameters of \texttt{ap\_rvputs} as variables
with \textsf{high} integrity, while the returned HTTP\_REFERER has
\textsf{low} integrity. The dual verification problem on integrity
means to decide the \textsf{high} variables hold the following
relation $(\upmu_1 =_H \upmu_2)\Rightarrow
\textsf{G}(n_{final}^*\Rightarrow (\upmu_1 =_H \upmu_2))$. Since
Moped can only deal with integer and boolean variables and array of
both types, we have to first abstract string to integer, and model
the string operations using binary operations of integers. To ensure
that the abstract integer variable could be assigned by constant
value modeling constant string, $N$ could not be less than 4. We
have successfully verified this program insecure, and according to
the experimental results, the efficiency of our method under proper
abstraction is competitive to other methods
\cite{DBLP:conf/pldi/UnnoKY06}.

\section{\label{sec:discuss}Discussions}
To the language presented in Section \ref{subsec:language}, we
require the \texttt{out} parameter of procedure to refer to global
variable. Because in symbolic pushdown rules of callee procedure,
the stack symbols do not contain the stack symbol of caller. We
cannot express the effect to the local variables of caller directly
in one pushdown rule, including \texttt{return} action, of callee.
This problem comes from the definition of transition relation
$\twoheadrightarrow $ of $\mathcal {I}_{\mathcal {P}}$ corresponding
to pushdown system $\mathcal {P}$. We can solve this problem by
storing the \texttt{out} value referring to the caller's local
variable in the common part of caller's rules and callee's rules,
that is the global variables as control locations. That means we
need to add an additional control location \texttt{r} for the
\texttt{out} value. In \texttt{P-PROC} of Fig.~\ref{fig:derivation},
we need to substitute \texttt{$x_2$} in procedure body \texttt{S}
with \texttt{r} instead of \texttt{l}, and add a rule for the caller
to assign value of \texttt{r} to its local variable. This additional
global variable could also help us to store the return value of
language whose procedure has side effect. A possible ambiguity is
the security level of this additional global variable. We could set
it \textsf{high} to avoid invalid information flow caused by its
change. Also we do not consider parameters with \textsf{inout} type
since an \textsf{inout} parameter can be trivially eliminated by one
\textsf{in} and one \textsf{out} parameter.

Model checking based approaches are usually argued against the
efficiency problem and state explosion. Although the derived forms
of self-composition we develop could greatly improve the efficiency,
the complexity of satisfiability for LTL is complete for
PSPACE\cite{DBLP:journals/jacm/SistlaC85}\cite{DBLP:conf/banff/Vardi95}.
Also the ordinary partial correctness specified by self-composition
is undecidable while type-based approaches are generally much
faster.

As we have stated in last section, Moped can only deal with boolean
and integer variables and arrays, and each integer variable should
have a finite range of $0\sim 2^N-1$. This makes abstraction
necessary for verifying realistic programs. So we plan to scale up
our approach to more real applications by adapting proper approaches
on abstraction, possibly using some recently developed
tools\cite{DBLP:conf/tacas/SuwimonteerabuthSE05}\cite{DBLP:conf/cav/SuwimonteerabuthBSE07}.
Because all forms of self-composition are developed on symbolic
pushdown system, the influence of abstraction on self-composition
should be little.

When the program has a great many constants, certain information
flow will be omitted if the initial $N$ is too small. But larger $N$
means increase on time and memory consumption, as we can see in
Fig.~\ref{fig:hu_result}. Therefore it is sometimes difficult to
balance the exactness and the efficiency by choosing proper initial
value of $N$. Another restriction of our approach is that with the
deterministic nature of the model our approach has not scaled to
concurrent programs. Our future work will adapt our approach to
handle concurrency using some formal
models\cite{DBLP:conf/popl/BouajjaniET03}\cite{DBLP:conf/fsttcs/BouajjaniESS05}.

\nocite{mcps:mcbook}
\bibliographystyle{splncs}
\bibliography{latex8}

\section*{Appendix}
\subsection*{Proof of Theorem~\ref{theorem:soundness}}
{\footnotesize

We give some preliminary definitions before the detailed proof. We
define a mapping $G:\mathbb{P}\times \mathbb{Q}\times \mathbb{I}
\rightarrow \mathbb{L}$ from local variables of pushdown system to
memory locations of current stack in operational semantics.
$\mathbb{I}$ contains the depth of procedure call from $main$ to the
procedure which local variable belongs to. If we have
$l=G(p_k,x,i)\in dom(\lambda)$, that means $l$ stands for local
variable $x$ of procedure $p_k$, and $p_k$ is on the $i$th position
of procedure stack. Let $i_t\in \mathbb{I}$ be the top-most
procedure of current stack. We claim a common constraint to the
language semantics that currently executing procedure cannot operate
on the local variables of other procedures. Thus if
$l\mathrel{\mathop:}= e\ (l\in dom(\lambda))$ is executed in calling
procedure $p_k$, there must be a $\langle p_k,x\rangle $, where
$l=G(p_k,x,i_t)$ and the corresponding statement is
$x\mathrel{\mathop:}= e$.

We inductively prove that for each derivation rule
$\Phi(S,n_i,n_j,p,\emptyset )$, the transition from $\upmu_i$ to
$\upmu_j$ is operationally sound to the change from
$(\mu_0,\lambda_0)$ to $(\mu',\lambda')$ if
$(\mu_0,\lambda_0,S)\downarrow (\mu',\lambda')$.

\begin{enumerate}[1.]
\item \textbf{SKIP}

$\mu'=\mu_0$, $\lambda'=\lambda_0$. (SKIP)\qquad $\rho'=\rho_0$, $\eta'(p)=\eta_0(p)$. (\texttt{P-SKIP})\\
$\forall l\in dom(\mu), \mu_0(l)=\rho_0[l]\Rightarrow \mu'(l)=\rho'[l]$\\
Since $\lambda'=\lambda_0$, $\forall l\in dom(\lambda')\Rightarrow
l\in dom(\lambda_0)$. If $l=G(p,x,i_t)$,
$\lambda_0(l)=\eta_0(p)[x]$, thus $\lambda'(l)=\eta_0(p)[x]$. Since
$\eta'(p)=\eta_0(p)$, we have $\forall x\in \eta(p),
\eta'(p)[x]=\eta_0(p)[x]$. Thus $\lambda'(l)=\eta'(p)[x]$.
\item \textbf{UPDATE}\\
If $l_0\in dom(\mu). (l_0\in \rho)$.\\
$\mu'=\mu_0[l_0\mathrel{\mathop:}= v],
\lambda'=\lambda_0$.(UPDATE-HEAP)\\
$\rho'=\rho_0[l_0\mathrel{\mathop:}= e],\eta'(p)=\eta_0(p)$.
(\texttt{P-UPDATE})\\
$\forall l\in dom(\mu)\setminus\{l_0\}, \mu'(l)=\mu_0(l),
\rho'[l]=\rho_0[l]$, since $\mu_0(l)=\rho_0[l]$, then
$\mu'(l)=\rho'[l]$, and since $\mu'(l_0)=\rho'[l_0]=v$, we have
$\forall l\in dom(\mu), \mu'(l)=\rho'[l]$.\\
Since $\lambda'=\lambda_0, \forall l\in dom(\lambda')\Rightarrow
l\in dom(\lambda_0)$. If $l=G(p,x,i_t), \lambda_0(l)=\eta_0(p)[x]$,
thus $\lambda'(l)=\eta_0(p)[x]$. Since $\eta'(p)=\eta_0(p)$, we have
$\forall x\in \eta(p), \eta'(p)[x]=\eta_0(p)[x]$. Thus
$\lambda'(l)=\eta'(p)[x]$.\\
If $l_0\in dom(\lambda_0), \exists \langle
p,x_0\rangle,l_0=G(p,x_0,i_t)$, the corresponding statement of
pushdown system is $x_0 \mathrel{\mathop:}= e$.\\
$\mu'=\mu_0, \lambda'=\lambda_0[l_0\mathrel{\mathop:}= v]$.
(UPDATE-STACK)\\
$\rho'=\rho_0, \eta'(p)=\eta_0(p)[x_0\mathrel{\mathop:}= e]$.
(\texttt{P-UPDATE})\\
$\forall l\in dom(\mu),\mu'(l)=\mu(l),\rho'[l]=\rho_0[l]$, since
$\mu_0(l)=\rho_0[l]$, then $\mu'(l)=\rho'[l]$.\\
$\forall l\in dom(\lambda')\setminus\{l_0\},
\lambda'(l)=\lambda_0(l)$, if $l=G(p,x,i_t),
\lambda_0(l)=\eta_0(p)[x] \wedge x \neq x_0$, thus we have
$\eta'(p)[x]=\eta_0(p)[x],
\lambda'(l)=\lambda_0(l)=\eta_0(p)[x]=\eta'(p)[x]$. Also we have
$\lambda'(l_0)=\eta'(p)[x_0]=v$. Therefore $\forall l\in
dom(\lambda')$, if $l=G(p,x,i_t),\lambda'(l)=\eta'(p)[x]$.

\item \textbf{SEQ}

$\left( \begin{array}{c} \left( \begin{array}{l} \forall l\in
dom(\mu_1),\mu_1(l)=\rho_1[l] \wedge \\\forall l\in dom(\lambda_1)$,
if $\exists \langle p,x\rangle,l=G(p,x,i_t)$, then
$\lambda_1(l)=\eta_1(p)[x]\end{array}
\right)\Rightarrow \\
\left( \begin{array}{l} \forall l\in dom(\mu_1),\mu_1'(l)=\rho_1'[l]
\wedge \\\forall l\in dom(\lambda_1')$, if $\exists \langle
p,x\rangle,l=G(p,x,i_t)$, then
$\lambda_1'(l)=\eta_1'(p)[x]\end{array} \right)\end{array}\right)$
\hfill (I)\\
$\left( \begin{array}{c} \left( \begin{array}{l} \forall l\in
dom(\mu_2),\mu_2(l)=\rho_2[l] \wedge \\\forall l\in dom(\lambda_2)$,
if $\exists \langle p,x\rangle,l=G(p,x,i_t)$, then
$\lambda_2(l)=\eta_2(p)[x]\end{array}
\right)\Rightarrow \\
\left( \begin{array}{l} \forall l\in dom(\mu_2),\mu_2'(l)=\rho_2'[l]
\wedge \\\forall l\in dom(\lambda_2')$, if $\exists \langle
p,x\rangle,l=G(p,x,i_t)$, then
$\lambda_2'(l)=\eta_2'(p)[x]\end{array} \right)\end{array}\right)$
\hfill (II)\\
$\left( \begin{array}{l} \forall l\in dom(\mu_0),\mu_0(l)=\rho_0[l]
\wedge \\\forall l\in dom(\lambda_0)$, if $\exists \langle
p,x\rangle,l=G(p,x,i_t)$, then $\lambda_0(l)=\eta_0(p)[x]\end{array}
\right)\wedge$\\
$\mu_1=\mu_0 \wedge \lambda_1=\lambda_0 \wedge \rho_1=\rho_0 \wedge
\eta_1(p)=\eta_0(p)\wedge $ (I) $\Rightarrow$\\
$\left( \begin{array}{l} \forall l\in
dom(\mu_1),\mu_1'(l)=\rho_1'[l] \wedge \\\forall l\in
dom(\lambda_1')$, if $\exists \langle p,x\rangle,l=G(p,x,i_t)$, then
$\lambda_1'(l)=\eta_1'(p)[x]\end{array} \right)$\hfill (III)\\
$\mu_2=\mu_1' \wedge \lambda_2=\lambda_1' \wedge \rho_2=\rho_1'
\wedge \eta_2(p)=\eta_1'(p) \wedge$ (III) $\Rightarrow$\\
$\left( \begin{array}{l} \forall l\in
dom(\mu_2),\mu_2'(l)=\rho_2'[l] \wedge \\\forall l\in
dom(\lambda_2')$, if $\exists \langle p,x\rangle,l= G(p,x,i_t)$,
then $\lambda_2'(l)=\eta_2'(p)[x]\end{array} \right)$\hfill (IV)\\
$\mu'=\mu_2' \wedge \lambda'=\lambda_2' \wedge \rho'=\rho_2' \wedge
\eta'(p)=\eta_2'(p) \wedge$ (IV) $\Rightarrow$\\
$\left( \begin{array}{l} \forall l\in dom(\mu'),\mu'(l)=\rho'[l]
\wedge \\\forall l\in dom(\lambda')$, if $\exists \langle
p,x\rangle,l=G(p,x,i_t)$, then $\lambda'(l)=\eta'(p)[x]\end{array}
\right)$.

\item \textbf{CONDITIONAL-BRANCH}

If $(\mu_0,\lambda_0,e)\downarrow true$, the branch conditions in
$R$ show that the transition from $n_i$ to $n_j$ is executed by
$\{\langle (\rho_i)(n_i,\eta_i(p))\rangle \hookrightarrow \langle
(\rho_i)(n_k,\eta_i(p))\rangle \mid R\cup \{e\}\}$ and the rules
generated by $\Phi(S_1,n_k,n_j,p,R\cup \{e\})$. We have
$\rho_k=\rho_0 \wedge \eta_k(p)=\eta_0(p)$,\\
$\left(\begin{array}{l} \forall l\in dom(\mu_0),
\mu_0(l)=\rho_0[l]=\rho_k[l] \wedge \\\forall l\in dom(\lambda_0)$,
if $\exists \langle p,x\rangle,l=G(p,x,i_t)$, then
$\lambda_0(l)=\eta_0(p)[x]=\eta_k(p)[x]
\end{array}\right)$\\
We can inductively get\\
$\left(\begin{array}{l} \forall l\in dom(\mu'), \mu'(l)=\rho'[l]
\wedge \\\forall l\in dom(\lambda')$, if $\exists \langle
p,x\rangle,l=G(p,x,i_t)$, then $\lambda'(l)=\eta'(p)[x]
\end{array}\right)$\\
If $(\mu_0,\lambda_0,e)\downarrow false$, the proof is similar.

\item \textbf{LOOP}

If $(\mu_0,\lambda_0,e)\downarrow false$,
$\mu'=\mu_0,\lambda'=\lambda_0$. (WHILE-F)\\
From the branch condition in $R$ of (\texttt{P-LOOP}) we can see the
transition is executed by pushdown rule $\{\langle
(\rho_i)(n_i,\eta_i(p))\rangle \hookrightarrow \langle
(\rho_i)(n_j,\eta_i(p))\rangle \mid R\cup \{!e\}\}$. Thus
$\rho'=\rho_0, \eta'(p)=\eta_0(p)$.\\
$\forall l\in dom(\mu_0), \mu_0(l)=\rho_0[l]\Rightarrow
\mu'(l)=\rho'[l]$\\
Since $\lambda'=\lambda_0, \forall l\in dom(\lambda')\Rightarrow
l\in dom(\lambda_0)$. If $l=G(p,x,i_t), \lambda_0(l)=\eta_0(p)[x]$,
thus $\lambda'(l)=\eta_0(p)[x]$. Since $\eta'(p)=\eta_0(p)$, then
$\forall x\in \eta(p), \eta'(p)[x]=\eta_0(p)[x]$. Thus
$\lambda'(l)=\eta'(p)[x]$.\\
If $(\mu_0,\lambda_0,e)\downarrow true$, we need to prove the
execution of pushdown system before the next time evaluation of $e$
is sound to operational semantics.\\
$\rho_q=\rho_0, \eta_q(p)=\eta_0(p)$.(\texttt{P-LOOP})\\
$\left(\begin{array}{l} \forall l\in dom(\mu_0),
\mu_0(l)=\rho_0[l]=\rho_q[l] \wedge \\\forall l\in dom(\lambda_0)$,
if $\exists \langle p,x\rangle,l=G(p,x,i_t)$, then
$\lambda_0(l)=\eta_0(p)[x]=\eta_q(p)[x]
\end{array}\right)$\\
$\Rightarrow \left(\begin{array}{l} \forall l\in dom(\mu'),
\mu'(l)=\rho_i'[l] \wedge \\\forall l\in dom(\lambda')$, if $\exists
\langle p,x\rangle,l=G(p,x,i_t)$, then $\lambda'(l)=\eta_i'(p)[x]
\end{array}\right)$.

\item \textbf{BINDVAR}

$\left(\begin{array}{l} \forall l\in dom(\mu_0), \mu_0(l)=\rho_0[l]
\wedge \\\forall l\in dom(\lambda_0)$, if $\exists \langle
p,x\rangle,l=G(p,x,i_t)$, then $\lambda_0(l)=\eta_0(p)[x]
\end{array}\right)$\hfill (I)\\
$\left( \begin{array}{c} \left(
\begin{array}{l} \forall l\in dom(\mu_1),\mu_1(l)=\rho_1[l] \wedge
\\\forall l\in dom(\lambda_1)$, if $\exists \langle
p,x\rangle,l=G(p,x,i_t)$, then $\lambda_1(l)=\eta_1(p)[x]\end{array}
\right)\Rightarrow\\
\left( \begin{array}{l} \forall l\in dom(\mu_2),\mu_2(l)=\rho_2[l]
\wedge \\\forall l\in dom(\lambda_2)$, if $\exists \langle
p,x\rangle,l=G(p,x,i_t)$, then $\lambda_2(l)=\eta_2(p)[x]\end{array}
\right)\end{array}\right)$\hfill(II)\\
$\mu_1=\mu_0, \lambda_1=\lambda_0\uplus [l_0\mathrel{\mathop:}=v]\
(l_0\notin dom(\mu_0) \wedge l_0\notin dom(\lambda_0))$. (BINDVAR)\\
$\rho_1=\rho_0, \eta_1(p)=\eta_0(p)[x_0\mathrel{\mathop:}=e]$.
(\texttt{P-BINDVAR})\\
Therefore, $\forall l\in dom(\mu), \mu_1(l)=\rho_1[l]$.\hfill(III)\\
$\forall l\in dom(\lambda_1)\setminus\{l_0\}, l\in dom(\lambda_0)
\wedge \lambda_0(l)=\lambda_1(l)$, we have, if $\exists \langle
p,x\rangle, l=G(p,x,i_t)$, then $\lambda_0(l)=\eta_0(p)[x] \wedge
x\neq x_0$, thus $\eta_1(p)[x]=\eta_0(p)[x]$.\\
Therefore $\forall l\in dom(\lambda_1)\setminus\{l_0\}$, if $\exists
\langle p,x\rangle,l=G(p,x,i_t), \lambda_1(l)=\eta_1(p)[x]$. And
since $\lambda_1(l_0)=\eta_1(p)[x_0]=v$, we have $\forall l\in
dom(\lambda_1)$, if $\exists \langle p,x\rangle,l=G(p,x,i_t)$, then
$\lambda_1(l)=\eta_1(p)[x]$.\hfill(IV)\\
From (III)(IV)(II) we have\\
$\left(\begin{array}{l} \forall l\in dom(\mu_2), \mu_2(l)=\rho_2[l]
\wedge \\\forall l\in dom(\lambda_2)$, if $\exists \langle
p,x\rangle,l=G(p,x,i_t)$, then $\lambda_2(l)=\eta_2(p)[x]
\end{array}\right)$\\
$\mu'=\mu_2, \lambda'=\lambda_2\setminus \{l_0\}$. (BINDVAR)\qquad $\rho'=\rho_2, \eta'(p)=\eta_2(p)$. (\texttt{P-BINDVAR})\\
Therefore, $ \left(\begin{array}{l} \forall l\in dom(\mu'),
\mu'(l)=\rho'[l] \wedge \\\forall l\in dom(\lambda')$, if $\exists
\langle p,x\rangle,l=G(p,x,i_t)$, then $\lambda'(l)=\eta'(p)[x]
\end{array}\right)$

\item \textbf{PROC}\\
$\left(\begin{array}{l} \forall l\in dom(\mu_0), \mu_0(l)=\rho_0[l]
\wedge \\\forall l\in dom(\lambda_0)$, if $\exists \langle
p,x\rangle,l=G(p,x,i_t)$, then $\lambda_0(l)=\eta_0(p)[x]
\end{array}\right)$\hfill (I)\\
$\left( \begin{array}{c} \left(
\begin{array}{l} \forall l\in dom(\mu_1),\mu_1(l)=\rho_1[l] \wedge
\\\forall l\in dom(\lambda_1)$, if $\exists \langle
p',x\rangle,l=G(p',x,i_t)$, then
$\lambda_1(l)=\eta_1(p')[x]\end{array}
\right)\Rightarrow \\
\left( \begin{array}{l} \forall l\in dom(\mu_2),\mu_2(l)=\rho_2[l]
\wedge \\\forall l\in dom(\lambda_2)$, if $\exists \langle
p',x\rangle,l=G(p',x,i_t)$, then
$\lambda_2(l)=\eta_2(p')[x]\end{array}
\right)\end{array}\right)$\hfill (II)\\
$\mu_1=\mu_0, \lambda_1=\lambda_0\uplus [l'\mathrel{\mathop:}=v]\
(l'\notin dom(\mu_0) \wedge l'\notin dom(\lambda_0))$. (CALL)\\
$\rho_1=\rho_0$. Therefore $\forall l\in dom(\mu_1),
\mu_1(l)=\rho_1[l]$.\hfill (III)\\
$\forall l\in dom(\lambda_1)\setminus\{l'\}, l\in dom(\lambda_0),
\neg \exists \langle p',x\rangle (l=G(p',x,i_t))$.\\
$l'=G(p',x_1,i_t) \wedge \lambda_1(l')=\eta_1(p')[x_1]=v$.\\
Therefore, $\forall l\in dom(\lambda_1), if\ \exists \langle
p',x\rangle,l=G(p',x,i_t), then\ \lambda_1(l)=\eta_1(p')[x]$.\hfill (IV)\\
From (III)(IV)(II) we have\\
$\left(\begin{array}{l} \forall l\in dom(\mu_2), \mu_2(l)=\rho_2[l]
\wedge \\\forall l\in dom(\lambda_2)$, if $\exists \langle
p',x\rangle,l=G(p',x,i_t)$, then $\lambda_2(l)=\eta_2(p')[x]
\end{array}\right)$\\
$\mu'=\mu_2,\lambda'=\lambda_2\setminus \{l'\}$. (CALL)\qquad
$\rho'=\rho_2,\eta'(p)=\eta_0(p)$. (\texttt{P-PROC})\\
Therefore, $\forall l\in dom(\mu), \mu'(l)=\rho'[l]$.\\
Procedure $p$ is on top of procedure stack, so
$dom(\lambda_0)=dom(\lambda')$. Since procedure $p'$ cannot modify
local variables in $\lambda'$, we have $\forall l\in dom(\lambda'),
\lambda'(l)=\lambda_0(l)$. Since $\eta'(p)=\eta_0(p)$, we have
$\forall l\in dom(\lambda')$, if $\exists \langle
p,x\rangle,l=G(p,x,i_t)$, then $\lambda'(l)=\eta'(p)[x]$.

\end{enumerate}

}

\end{document}